\documentclass[conference]{IEEEtran}

\pdfoutput=1

\newcommand{\CC}{\mathbb{C}}
\newcommand{\RR}{\mathbb{R}}

\newcommand{\eu}  {\mathrm{e}}

\renewcommand{\Pr} {\mathop\mathrm{Pr}\nolimits}

\newcommand{\E}{\mathbb{E}}
\newcommand{\reff}[1]{\eqref{#1}}
\newcommand{\defeq}{\triangleq}

\newcommand{\av}{\boldsymbol{a}}
\newcommand{\bv}{\boldsymbol{b}}

\newcommand{\Hm}{\boldsymbol{H}}
\newcommand{\Id}{\boldsymbol{I}}

\newcommand{\Cc}{\mathcal{C}}

\newcommand{\Nc}{\mathcal{N}}
\newcommand{\Oc}{\mathcal{O}}

\newcommand{\Rc}{\mathcal{R}}
\newcommand{\Sc}{\mathcal{S}}

\newcommand{\Xc}{\mathcal{X}}
\newcommand{\Yc}{\mathcal{Y}}

\newcommand{\rhov}{\boldsymbol{\rho}}

\newcommand{\Sigmam}{\boldsymbol{\Sigma}}

\usepackage{cite}      

\usepackage{graphicx}  

\usepackage{psfrag}    

\usepackage{subfigure} 

\usepackage{url}       

\usepackage{stfloats}  

\usepackage{amsmath}   
\usepackage{array}
 \makeatletter
 \let\NAT@parse\undefined
 \makeatother
\usepackage{times}
\usepackage{epsfig,graphicx,color,pstricks}
\usepackage{amsmath,amssymb,amsbsy,amsthm,amsfonts,latexsym}
\usepackage{bm}
\usepackage{psfrag}
\usepackage[mathscr]{eucal}
\usepackage{cite}
\usepackage[all]{xy}
\usepackage{subfigure}
\usepackage{graphics}

\newtheorem{theorem}{Theorem}

\newtheorem{corollary}[theorem]{Corollary}

\begin{document}

\title{$K$-user Interference Channels: General Outer Bound and Sum-capacity for Certain Gaussian Channels}

\author{%
\authorblockN{Daniela Tuninetti}
\authorblockA{Department of Electrical and Computer Engineering,\\
University of Illinois at Chicago, Illinois 60607, USA,\\
Email: {\tt danielat@uic.edu} }
}
\maketitle

\begin{abstract}
This paper derives an outer bound on the capacity region
of a general memoryless interference channel with an arbitrary 
number of users.  The derivation follows from a generalization
of the techniques developed by Kramer and 
by Etkin {\em et al} for the Gaussian two-user channel. 
The derived bound is the first known outer bound 
valid for {\em any memoryless channel}.

In Gaussian noise, classes of channels 
for which the proposed bound gives the sum-rate capacity
are identified, including  
degraded channels and a class of Z-channels. 
\end{abstract}


\section{Introduction}
\label{sec:intro}

\begin{figure}
	\centering
		\includegraphics[width=7cm]{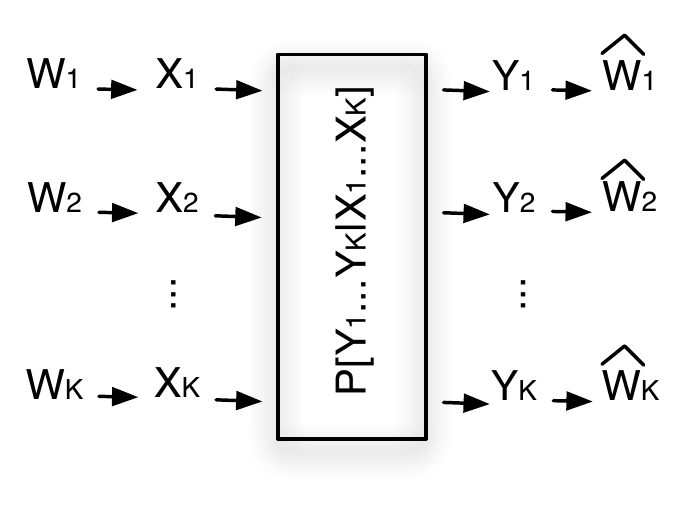}
	\vspace*{-0.5cm}
	\caption{The general memoryless InterFerence channel with $K$ source-destination pairs ($K$-IFC) considered in this work.}
	\label{fig:DMCchannel}
\end{figure}

Determining the ultimate capacity limits of the general memoryless 
$K$-user InterFerence Channel ($K$-IFC) is an open problem
since its inception. 
The network considered in this work is depicted in Fig.~\ref{fig:DMCchannel}:
it consists of $K$ pairs of nodes and is defined by input alphabets $(\Xc_1, \ldots, \Xc_K)$,
output alphabets $(\Yc_1, \ldots, \Yc_K)$, and a 
channel transition probability $P_{Y_1,\ldots,Y_K|X_1,\ldots,X_K}$.
The only assumption on the channel is that it is memoryless.
Source $i$, $i\in[1:K]$, wishes to communicate to destination $i$
an independent message $W_i$. A $(\eu^{n R_1},\ldots, \eu^{n R_K}, n, \epsilon_n)$ code
consists of $K$ encoding functions
$[1:\eu^{n R_i}] \to \Xc_{i}^{n}$,
$K$ decoding functions
$\widehat{W_i}: \Yc_{i}^{n}  \to [1:\eu^{n R_i}]$,
such that
$\Pr\big[\widehat{W_i}(Y_{i}^n)\not= W_i\big] \leq  \epsilon_n$, $i \in[1:K]$.
A rate-tuplet $(R_1,\ldots,R_K)$ is achievable if there exists a family
of $(\eu^{n R_1},\ldots, \eu^{n R_K}, n, \epsilon_n)$ codes such that
$\epsilon_n\to 0$ as $n\to \infty$.
The capacity region is the convex closure
of the set of achievable rate-tuplet $(R_1,\ldots,R_K)$.
The capacity is not known in general.

{\em The goal of this paper is to derive an outer bound for the $K$-IFC
that holds for any memoryless channel (not necessarily Gaussian)
and for any $K\geq 2$.}

\subsection{Past Work}
\label{subsec:past}

The capacity region of the 2-IFC
is known 
if the interference is strong~\cite{sato:IFCoutit1978,costa_elgamal:strong:it1987,carleial:ifcstrong:it1975},
if the channel outputs are deterministic and invertible functions of the inputs~\cite{costa_aelgamal:it1982}, and
if the channel has a special form of degradeness~\cite{liu_ulukus:degifc,benzel:it1979}.
The largest known inner bound is due to Han and Kobayashi (HK)~\cite{Han_Kobayashi:it1981} and uses
rate splitting and joint decoding.  General outer bounds are due to Sato~\cite{sato:IFCoutit1977,sato:IFCoutit1978},
and Carleial~\cite{carleial:IFCit1983} (see also Kramer~\cite[Th.5]{kramer:it2004}).

For the Gaussian 2-IFC, the capacity region is fully known
in strong interference only~\cite{carleial:ifcstrong:it1975,sato:IFCstrong:it1981,costa:awgnifc:it1985}.
The sum-capacity is however known
in mixed interference~\cite{tuninettiweng:isit2008,canadanoisy},
for the Z-channel~\cite{sason:it2004}, and
in very weak interference~\cite{biaonoisy2008:IT09,venunoisy2008:IT09,canadanoisy}. 
In mixed and weak interference, a simple rate splitting in the HK region is optimal
to within one~bit~\cite{etkin_tse_hua:withinonebit:subIt06}.
The best outer bound may be obtained by intersecting the regions
derived by Kramer in~\cite{kramer:it2004},
by Etkin {\em et al.} in~\cite{etkin_tse_hua:withinonebit:subIt06}, and the region independently obtained in~\cite{biaonoisy2008:IT09,venunoisy2008:IT09,canadanoisy} and later tighten by Etkin in~\cite{EtkinISIT2009}.

Few results are available for more than two users and/or for non-Gaussian channels.
Please refer to~\cite{tuniITA11:G3IFCsumrate}, and references therein,
for a more detailed discussion of the
past work that we shall only briefly list in the following for sake of space.

General inner bound regions are lacking.
A straightforward generalization of the HK approach, 
whereby each user has a different message for every subset of 
non-intended receivers, 
has a super-exponential complexity
in the number of users 
and might be suboptimal in general.
In fact, 
coding schemes that deal directly with the effect of the aggregate
interference, rather than with each interferer separately, as with
interference alignment~\cite{Cadambe-Jafar:arXiv:0707.0323v2}
and with structured codes~\cite{Nazer:2007}, are known
to achieve a larger number of degrees of freedom than simple HK schemes
for the Gaussian noise channel~\cite{Cadambe-Jafar:arXiv:0707.0323v2,Bresler:1toM_Mto1:allerton07,Sridharan:layredlattice:allerton08}.
To the best of the author's knowledge, no outer bounds have been
developed for the general (i.e., non-Gaussian) IFC with more than two users.

In Gaussian noise, channels with a special structure have been investigated such as:
%
the ``fully symmetric'' channel~\cite{cadambe:notseparable:it09,josesriram:kifcmacbound:itw10,biaonoisy2008:IT09,venunoisy2008:IT09},
%
the ``cyclic symmetric'' channel\cite{ZhouYu:cyclicsym:CISS10,Bandemer:cyclicsym:isit09},
%
the three-user channel with ``cyclic mixed strong-very strong'' interference~\cite{segzin_3ifc:arXiv:1010.4911},
%
and the ``one-to-many'' 
and the ``many-to-one'' 
channel~\cite{Bresler:1toM_Mto1:allerton07,Jovicic_Wang_Viswanath:itw2007,venunoisy2008:IT09}.
The Degrees of Freedom (DoF) of the Gaussian $K$-IFC has received more attention
~\cite{Sridharan:verystrongKifc:globecom08,jafar:dofKifc:it10,EtkinOrdentlichISIT2009,Sridharan:layredlattice:allerton08,Cadambe-Jafar:arXiv:0707.0323v2}; the lesson from the DoF analysis is that structured codes appear to outperform purely random codes and that the high-SNR
analysis is very sensitive to the way the $K^2$ parameters of the $K$-IFC are let grow to infinity.

Of direct relevance for this work are the Gaussian 2-IFC outer bounds 
derived by Kramer in~\cite[Th.1]{kramer:it2004} and
by Etkin {\em et al.} in~\cite{etkin_tse_hua:withinonebit:subIt06},
which we seek to generalize to any memoryless channel with any number 
of users.  The basic idea is to give side information to the receiver(s)
in such a way that the resulting bound can be single-letterized, does not
involve auxiliary random variables and can be computed for channels of
interest, such as the Gaussian channel.
An extension of~\cite[Th.1, first proof]{kramer:it2004} to the $K$-user Gaussian channel,
with any $K\geq 2$, appeared in~\cite{josesriram:kifcmacbound:itw10}; the idea is to provide a group of receivers
with sufficient side information so that they can decode a subset of the users as in a
Multiple Access Channel (MAC) channel--as also discussed in~\cite{cadambe:notseparable:it09};
the resulting optimization problem however does not appears to have a closed-form solution in general
(a closed form result was given in~\cite{josesriram:kifcmacbound:itw10} for degraded channels only)
and iterative algorithms for its numerical evaluation are discussed in~\cite{jose:icc2011}.
{\em In this work, we will approach the problem from a different angle: 
we generalize~\cite[Th.1, second LMMSE-based proof]{kramer:it2004}
rather than~\cite[Th.1, first ``general optimization problem''-based proof]{kramer:it2004}. 
We will show that our bound can be evaluated in closed-form for certain Gaussian channels
and it is sum-capacity for some classes of channels.}
Extensions of~\cite[Th.1]{etkin_tse_hua:withinonebit:subIt06} to the $K$-user Gaussian channel
appeared in~\cite{biaonoisy2008:ISIT08,venunoisy2008:IT09}. In both works,
the receivers are given a side information signal that generalizes that 
of~\cite[Th.1]{etkin_tse_hua:withinonebit:subIt06} whereby
entropy terms are related by using the entropy power (EPI)~\cite{book:cover_thomas:it}
and/or the extremal inequality (EI)~\cite{viswanathextrema}
rather than chosen so that they cancel one another.
{\em In this work we simply generalize the approach
of~\cite[Th.1]{etkin_tse_hua:withinonebit:subIt06} to any memoryless
channel as it is not obvious what EPI and/or EI are for a general channel.
}

\subsection{Contributions and Paper Organization}
The main contributions of this work are:
\begin{enumerate}
\item
In Section~\ref{sec:maindmc} we derive an outer bound on the
capacity region of the general memoryless IFC, i.e., not necessarily Gaussian,
with an arbitrary number of source-destination pairs.
\item
In Section~\ref{sec:maingauss} we specialize the bound derived in
Section~\ref{sec:maindmc} to the Gaussian channel. 
In~\cite{tuniITA11:G3IFCsumrate}, we showed that there exist channel parameters for which 
our proposed bound is the tightest known for the sum-rate of the Gaussian 3-IFC.
Here, we derive the sum-capacity of certain Z-like  Gaussian $K$-IFCs.
We also discuss how to generalize this sum-capacity result to non-Z 
Gaussian channels;
in particular we offer two alternative proofs for the
sum-capacity of the Gaussian degraded channel originally derived in~\cite{josesriram:kifcmacbound:itw10}.
\end{enumerate}
Section~\ref{sec:conc} concludes the paper.
Some of the proofs are in the Appendix.

\section{Main Result}\label{sec:maindmc}
\begin{figure}
	\centering
	\hspace*{-0.5cm}
		\includegraphics[width=5cm]{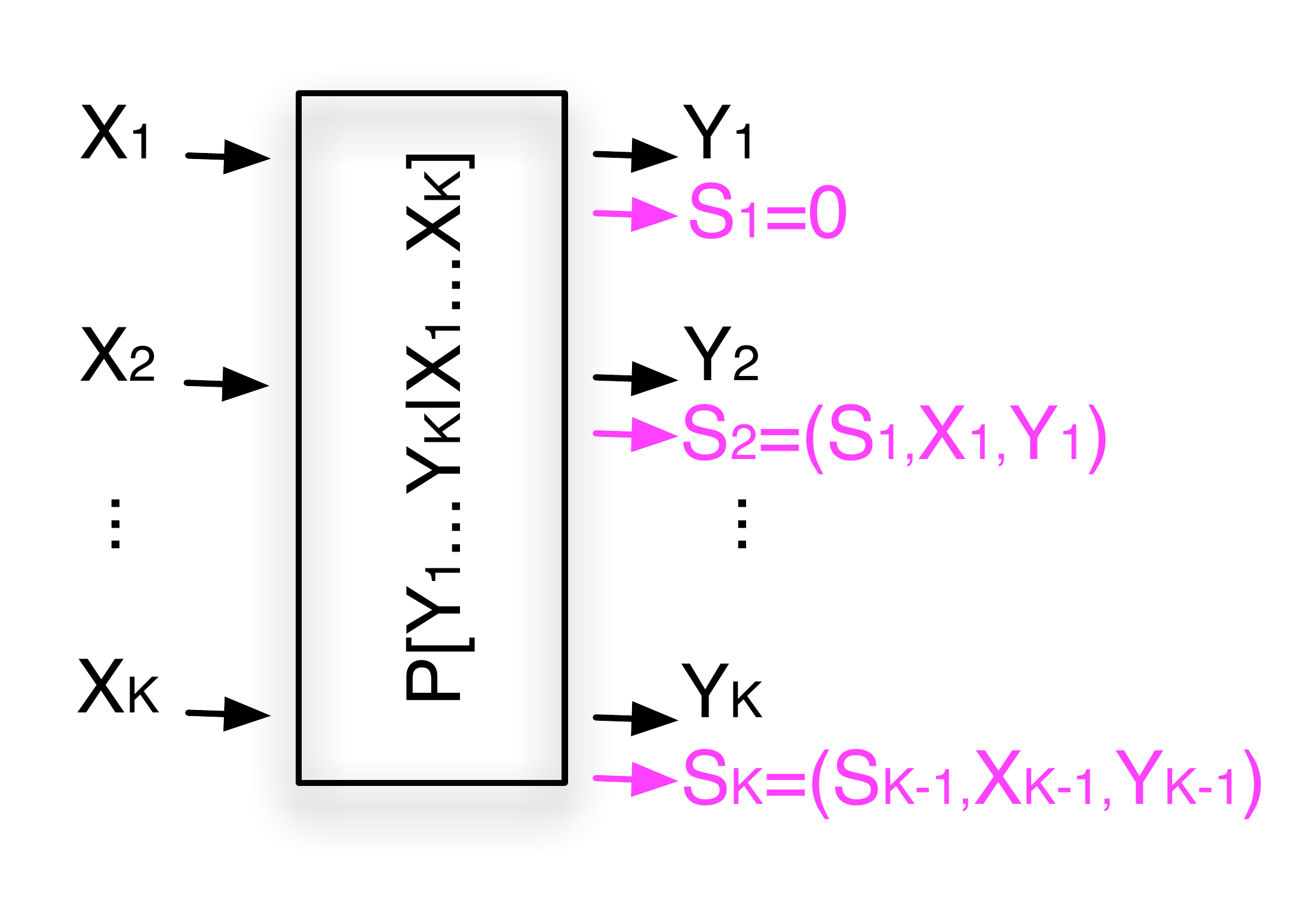}
	\hspace*{-0.4cm}
		\includegraphics[width=4cm]{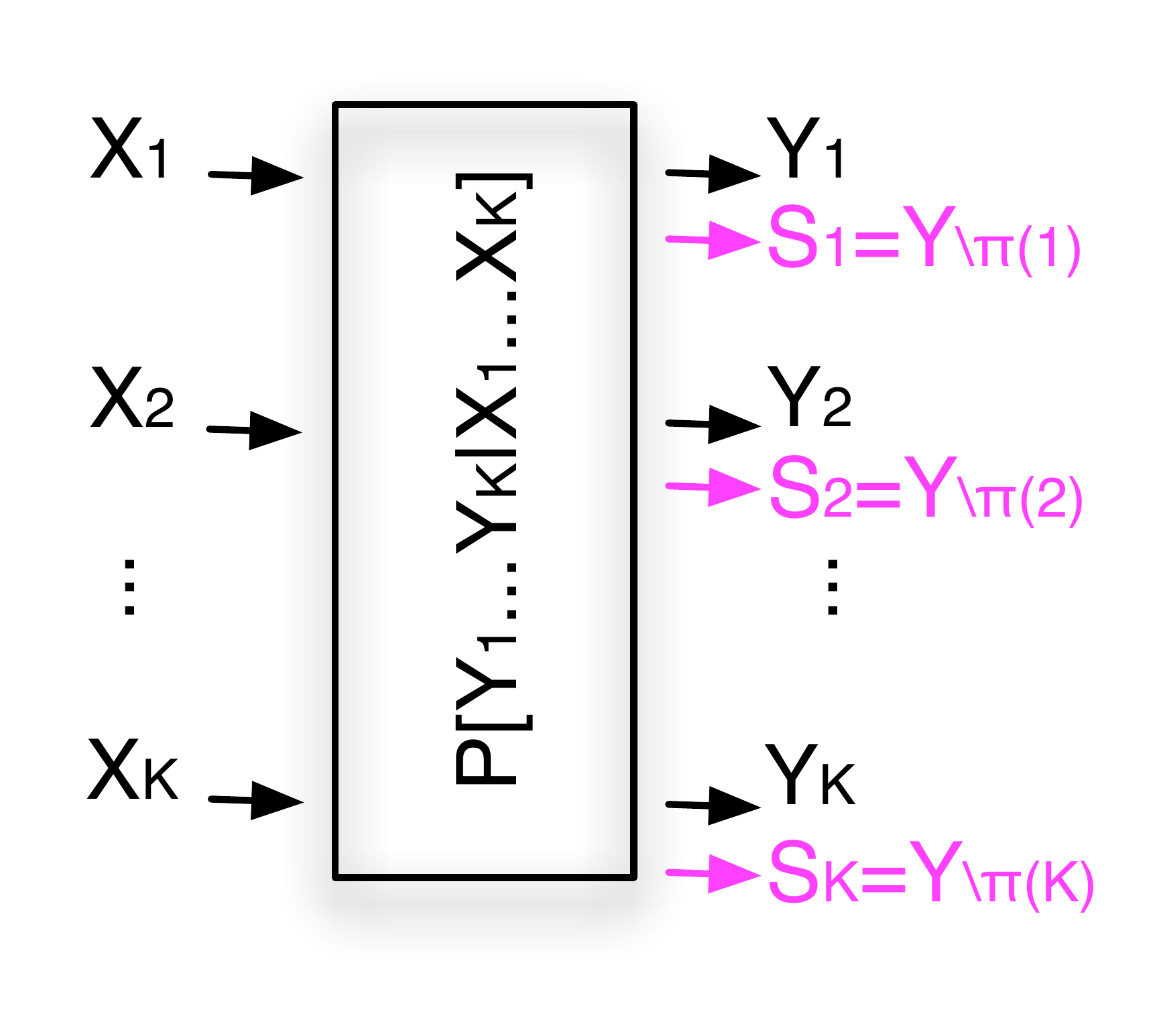}
	\vspace*{-0.5cm}
	\caption{Receiver side information for the proof of Th.\ref{th:out}.
	Left: for the bound in~\reff{eq:th:out kra} with $\Sc=[1:K]$ and $\pi=(1,\ldots,K)$.
	Right: for the bound in~\reff{eq:th:out etw}.}
	\label{fig:sideinfo}
\end{figure}

\begin{theorem}\label{th:out}
The capacity region of a general memoryless
$K$-IFC is contained into:
\begin{subequations}
\begin{align}
&\Oc_{\rm K-IFC}
= \bigcup_{P: P_{Q}\prod_{k=1}^{K}P_{X_k|Q} } \ 
\bigcap_{(\Sc,\pi): \Sc \subseteq [1:K], \pi \in \Pi[\Sc]}
\bigg\{
\nonumber\\&
\sum_{u\in \Sc}R_u \leq \sum_{k=1}^{|\Sc|} 
\nonumber\\&
I\Big( Y_{\pi_k};X([\pi_k:\pi_{|\Sc|}])\Big| 
       X([\pi_0:\pi_{k-1}]),Y([\pi_0:\pi_{k-1}]),
       X(\Sc^c),Q \Big),
\label{eq:th:out kra}
\\&
\sum_{u\in \Sc}R_u \leq \sum_{k=1}^{|\Sc|} 
H(Y_{k}|Y_{\backslash \pi_k})-
H(Y_{\backslash \pi_k}|Y_{k}, X_1,\ldots,X_K),
\label{eq:th:out etw}
\bigg\},
\end{align}
\label{eq:th:out}
\end{subequations}
where
the union is over all input distributions 
$P=P_{X_1,\ldots,X_K,Q}$ that factorize as
$P_{Q}\prod_{k=1}^{K}P_{X_k|Q}$ and the intersection is over all
subsets $\Sc$ of the user-index set $[1:K]$ and over all 
permutations $\pi$ of the elements of $\Sc$.
The random variable $Y_{\backslash k}$ has the same distribution of
$Y_k|_{X_k}$ (i.e., the conditional distribution of an output given its 
intended input).
A random variable with subscript $\pi_{0}$ is a constant.
\end{theorem}
\begin{IEEEproof}
The details of the proof may be found in the Appendix.
The key idea is to provide the $k$-th receiver, $k\in[1:K]$,
with the side information $S_k$ shown in Fig.~\ref{fig:sideinfo}. 
\end{IEEEproof}

\medskip
We remark that:
\begin{enumerate}

\item
Th.\ref{th:out} holds for {\em any} memoryless IFC
and for {\em any} number of users.

\item
Since the capacity region of a $K$-user IFC does not depend on the
joint transition probability $P_{Y_1,\ldots,Y_K|X_1,\ldots,X_K}$
(because the receivers cannot cooperate), but only on the
marginal transition probabilities $P_{Y_k|X_1,\ldots,X_K}$, $k\in[1:K]$,
each bound in Th.~\ref{th:out} (one for each pair $(\Sc,\pi)$)
can be optimized with respect to 
the joint probability $P_{Y_1,\ldots,Y_K|X_1,\ldots,X_K}$
as long as the marginal probabilities are preserved.

\item
The bound in~\reff{eq:th:out kra}
reduces to~\cite[Th.1]{kramer:it2004} for the
Gaussian 2-IFC when $\Xc_3=\emptyset$ (see~\cite[eq.(34)]{kramer:it2004}
which inspired the side information structure given on the
left side of Fig.~\ref{fig:sideinfo}).

\item
The bound in~\reff{eq:th:out etw} reduces to~\cite[Th.1]{etkin_tse_hua:withinonebit:subIt06} for the Gaussian 2-IFC
when $\Xc_3=\emptyset$ by setting $S_1 = Y_{\backslash 2}$
and  $S_2 = Y_{\backslash 1}$.
The bound in~\reff{eq:th:out etw} is tighter than~\cite[Th.1]{etkin_tse_hua:withinonebit:subIt06} 
because the correlation coefficient between the Gaussian noise of the channel
output $Y_k$ and the Gaussian noise of the side information $Y_{\backslash \pi_k}$,
$(k,\pi_k)\in[1:K]^2$, can be optimized so as to get the tightest bound.
It is however not tighter than the bound
independently obtained in~\cite{biaonoisy2008:IT09,venunoisy2008:IT09,canadanoisy} 
for the 2-user Gaussian channel.

\item
From~\reff{eq:th:out kra} we get
$N(K)=\sum_{k=1}^{K}{K \choose k} k!$ rate bounds.
For $K=2$, the $N(2)=4$ bounds are as in~\cite[Th.1]{kramer:it2004}
(two single-rate bounds and two sum-rate bounds).
For $K\geq3$, the $N(K)-K^2$ bounds that involve at least three rates
cannot be simply derived 
by silencing all but two users and then by applying~\cite[Th.1]{kramer:it2004} to the resulting 2-IFC.
The number of bounds grows exponentially with $K$:
$N(3)=15,$ 
$N(4)=52,$
$N(5)=325,$
etc.

Similarly, from~\reff{eq:th:out etw} we get $N(K)$ bounds;
those that involve at least three rates
cannot be obtained by simply applying the 2-IFC sum-rate
bound in~\cite[Th.1]{etkin_tse_hua:withinonebit:subIt06}.

\item
Th.\ref{th:out} can be easily evaluated. 
For example, the ``Gaussian maximizes entropy''~\cite{book:cover_thomas:it}
suffices to guarantee that a jointly Gaussian input is optimal
for Gaussian channels.

\item
Th.\ref{th:out} can be extended to other
memoryless channels without receiver cooperation. 
For example, 
the 2-user cognitive channel was considered in~\cite{RTDjournal1},
the 2-IFC with a cognitive relay in~\cite{rini2010CIFC+CR:ITWDublin},
and the 2-IFC with generalized feedback
(a.k.a. source cooperation) 
in~\cite{tuniISIT09out,tuniITA10out,vinodSC2009}.

For the $K$-IFC with generalized feedback for example, 
Th.\ref{th:out} must be modified as follows:
(a) replace each channel output $Y_k$ with the pair $(Y_k,Y_{{\rm GF,}k})$, where
$Y_{{\rm GF,}k}$ is the channel output observed at transmitter $k$, $k\in[1:K]$, 
(b) consider the union over all possible joint input distributions
$P_{X_1,\ldots,X_K}$
(because the generalized feedback enables source cooperation which results in
correlated inputs);
(c) choose the worst joint transition probability
$P_{Y_1,\ldots,Y_K|X_1,\ldots,X_K,Y_{{\rm GF,}1},\ldots,Y_{{\rm GF,}K}}$
that preserves the marginals
$P_{Y_k|X_1,\ldots,X_K,Y_{{\rm GF,}1},\ldots,Y_{{\rm GF,}K}}$, $k\in[1:K]$.

Similar extensions are possible for other channels.

\item
For the Gaussian 2-IFC,
besides the bounds in~\cite{kramer:it2004} and in~\cite{etkin_tse_hua:withinonebit:subIt06}
that we generalized in Th.\ref{th:out}, the following outer bounds are known:
~\cite[Th.2]{kramer:it2004}~\cite{biaonoisy2008:IT09,venunoisy2008:IT09,canadanoisy}
and~\cite{EtkinISIT2009}.  These bounds are tighter than~\cite[Th.1]{etkin_tse_hua:withinonebit:subIt06}
for some weak interference parameters.  
It is left for future work to generalize these 2-user Gaussian channel bounds
to non-Gaussian channels with more than two users.
We note that the common feature of these bounds is
to generalize the class of genie signals of~\cite[Th.1]{etkin_tse_hua:withinonebit:subIt06}
by relating entropy terms rather than canceling them
(see proof of Th.\ref{th:out}); this is done
by using the entropy power (EPI)~\cite{book:cover_thomas:it}
and/or the extremal inequality (EI)~\cite{viswanathextrema}; the extension
of the EPI and/or the EI to general (i.e., non-Gaussian) channels
is not trivial.

\end{enumerate}


\section{Gaussian channels}\label{sec:maingauss}
In this section we first introduce the Gaussian channel model
(subsection~\ref{sec:maingauss:mod}).
We then show that Th.\ref{th:out} eq.~\reff{eq:th:out kra}
gives the sum-capacity
for certain Z-channels (subsection~\ref{sec:maingauss:Z}). 
We conclude with Subsection~\ref{sec:maingauss:nonZ} where
we discuss how to extend the result of Subsection~\ref{sec:maingauss:Z}
to non-Z channels;
in doing so we show that Th.\ref{th:out} eq.~\reff{eq:th:out kra}
gives the sum-rate capacity of degraded channels, thereby
providing an alternative proof for 
the result of~\cite{josesriram:kifcmacbound:itw10};
we also offer an alternate proof
for the sum-rate capacity of the degraded Gaussian $K$-IFC 
by generalizing an argument originally devised by Sato 
for the degraded Gaussian 2-IFC~\cite{sato:IFCoutit1977}.

\subsection{The Gaussian Channel Model}\label{sec:maingauss:mod}
A SISO (single input single output) complex-valued
Gaussian $K$-IFC in {\em standard} form
has outputs:
\begin{align*}
     Y_{i} &= \sum_{k=1}^{K} h_{i,k}X_{k} + Z_{i}, 
\end{align*}
with input power constraint $\E[|X_{i}|^2] \leq 1$ and noise 
$Z_{i} \sim \Nc(0,1)$, $i \in[1:K]$.
The correlation among the Gaussian noises is irrelevant since the
capacity only depends on the marginal noise distributions.
The channel gains are fixed 
and are known to all terminals.
Without loss of generality, the direct link gains $h_{i,i}$, $i \in[1:K]$,
can be taken to be real-valued (because receiver $i$
can compensate for the phase of $h_{i,i}$)
and strictly positive (if $|h_{i,i}|^2=0$ then the SNR at receiver $i$ is
zero even in absence of interference, which implies that $R_i=0$ is optimal, i.e.,
the system has effectively one less user).
The Gaussian $K$-IFC is completely specified
by the channel matrix
$\Hm:  [\Hm]_{i,j}=h_{i,j}$, $(i,j) \in[1:K]\times[1:K]$.
 
In the following we adopted the Matlab-like convention that
$\Hm_{\Rc,\Cc}$ in the $|\Rc|\times|\Cc|$ matrix 
obtained from $\Hm$ by retaining the rows indexed by $\Rc$
and the columns indexed by $\Cc$.

\subsection{Sum-capacity of Z-like channels}\label{sec:maingauss:Z}

Here we consider a class of Gaussian $K$-IFCs for which
the channel matrix $\Hm$ is upper triangular.
This class of channels can be thought of 
as the multi-user generalization of the 2-IFC Z-channel~\cite{sason:it2004}.
The following theorem establishes the sum-capacity for
a subset of Z-channels for which treating interference as noise is
optimal:
\begin{theorem}\label{th:sum capacity of some Z channels}
Consider a $K\times K$ noise covariance matrix
$\Sigmam_{K}$ defined recursively as follows:
let $\Sigmam_{1}=[1]$ and $\forall k=2,\ldots,K$ let
\begin{align*}
  \Sigmam_{k}&= 
\begin{pmatrix}
\Sigmam_{k-1}   & \rhov_{k-1}\\
\rhov_{k-1}^H       & 1 \\
\end{pmatrix},
  \rhov_{k-1}\in\CC^{k-1}:
\rhov_{k-1}\rhov_{k-1}^H\preceq   \Sigmam_{1:k-1}.
\end{align*}
Consider a channel matrix $\Hm$ whose upper triangular part is
defined recursively as follows: for $k=K,\ldots,2$
\begin{align}
\Hm_{[1:k-1],[k]}
  &=\frac{\Hm_{[k],[k]}}{1+\| \Hm_{[k],[k+1;K]} \|^2} \nonumber
\\&\big(\rhov_{k-1} + \Hm_{[1:k-1],[k+1:K]} \Hm_{[k],[k+1:K]} \big),
\label{eq:channel for succ dec}
\end{align}
while the entries below the main diagonal of $\Hm$ are zero.
For the channel defined by~\reff{eq:channel for succ dec}, the sum-rate capacity is
given by~\reff{eq:th:out kra} and equals:
\begin{align}
\sum_{k=1}^{K} R_k
\leq
\sum_{k=1}^{K} \log\left(1+\frac{|h_{k,k}|^2}{1+\sum_{i=k+1}^{K}|h_{k,i}|^2}\right).
\label{eq:successive decoding sumrate}
\end{align}
\end{theorem}
\begin{IEEEproof}
Since every mutual information term in~\reff{eq:th:out kra} contains
all the inputs, the ``Gaussian maximizes entropy'' principle~\cite{book:cover_thomas:it}
assures that iid $\Nc(0,1)$ inputs are optimal. 
Consider $\Sc=[1:K]$ with $\pi=(1,\ldots,K)$ in~\reff{eq:th:out kra}
and rewrite the sum-rate as:
\begin{align}
&\sum_{k=1}^{K} R_k  \leq \min_{\Sigmam_{K}}
\left\{\sum_{k=1}^{K}I_k(\Sigmam_{k}) \right\},
\label{eq:th:out kra again}
\\&I_k(\Sigmam_{k}) \defeq I(Y_1,\ldots,Y_{k-1},Y_{k}; X_k|X_1,\ldots,X_{k-1}). 
\label{eq:th:out kra Ik}
\end{align}
The channel matrix $\Hm$ defined by~\reff{eq:channel for succ dec}
is such that for each $k=K,\ldots,2$:
\begin{align*}
  & \E[Y_1,\ldots,Y_{k-1}|Y_{k}, \ X_1,\ldots,X_{k-1}, \ X_k]
\\=&\E[Y_1,\ldots,Y_{k-1}|Y_{k}, \ X_1,\ldots,X_{k-1}],
\end{align*}
that is, conditioned on $(X_1,\ldots,X_{k-1})$ the set of outputs
$(Y_1,\ldots,Y_{k-1})$ is a degraded version of $Y_k$ and thus:
\begin{align}
I_k(\Sigmam_{k}) 
  &= I(Y_{k}; X_k|X_1,\ldots,X_{k-1})\nonumber
\\&= \log\left(1+\frac{|h_{k,k}|^2}{1+\sum_{i=k+1}^{K}|h_{k,i}|^2}\right)
\defeq r_{k}.
\label{eq:succ dec rates}
\end{align}
By summing the rates in~\reff{eq:succ dec rates}
over all $k\in[1:K]$ we obtain the sum-rate upper bound in~\reff{eq:successive decoding sumrate}.
The upper bound in~\reff{eq:successive decoding sumrate} can be achieved by simply
treating interference as noise at each receiver (recall that for the Z-channel,
the $k$-th receiver is interfered by $(X_{k+1},\ldots,X_{K})$ only).
\end{IEEEproof}

\smallskip
By considering all possible covariance matrices $\Sigmam_{K}$,
Th.\ref{th:sum capacity of some Z channels} identifies
a novel class of channels for which treating interference as noise
is sum-rate optimal (besides those in
\cite[Th.4, Th.5, and Th.7]{venunoisy2008:IT09} and~\cite[Th.3]{biaonoisy2008:ISIT08})
as shown in the following examples.
The correspondence between channel matrices and noise covariance matrices
given by~\reff{eq:channel for succ dec} is interesting in itself and deserves
further analysis.

\smallskip
{\bf Example~1.}
Th.\ref{th:sum capacity of some Z channels} 
assures that treating interference as noise is optimal for all
channels that can be built as in~\reff{eq:channel for succ dec}
from a covariance matrix of the type:
\begin{align*}
  \Sigmam_{K}&= 
\begin{pmatrix}
1     & v_2 & \ldots    & v_K \\
v_2^* &     &           &     \\
\vdots  &   & \Id_{k-1} &     \\
v_K^* &     &           &     \\
\end{pmatrix}:
\sum_{k=2}^{K}|v_k|^2 \leq 1.
\end{align*}
The resulting channel has gains $h_{1,k} = v_k h_{k,k}$, $k=2,\ldots,K$
and zero for the remaining non-diagonal entries; this channel
is to the so-called {\em many-to-one channel}~\cite{Bresler:1toM_Mto1:allerton07}.
The condition $\sum_{k=2}^{K}|v_k|^2 \leq 1$ identifies a subset of 
many-to-one channels for which treating interference as noise is optimal.
The condition $\sum_{k=2}^{K}|v_k|^2 \leq 1$ is equivalent to~\cite[Th.4]{venunoisy2008:IT09},
thus Th.\ref{th:sum capacity of some Z channels}  generalizes~\cite[Th.4]{venunoisy2008:IT09}.

\smallskip
The relationship between the class of channels identified by Th.\ref{th:sum capacity of some Z channels}
and that identified by~\cite[Th.3]{biaonoisy2008:ISIT08} (of which~\cite[Th.4 and Th.5]{venunoisy2008:IT09}
are special cases) is subject of current investigation.
We note that~\cite[Th.3]{biaonoisy2008:ISIT08}  is obtained from a generalization of~\cite[Th.1]{etkin_tse_hua:withinonebit:subIt06}
while Th.\ref{th:sum capacity of some Z channels} from a generalization of~\cite[Th.1]{kramer:it2004},
it is thus possible that~\cite[Th.3]{biaonoisy2008:ISIT08} and Th.\ref{th:sum capacity of some Z channels}
do not imply one another.



\smallskip
{\bf Example~2.}
Consider channels that can be obtained as in~\reff{eq:channel for succ dec}
from a rank-one covariance matrix of the type:
\begin{align*}
  \Sigmam_{K}: [\Sigmam_{K}]_{i,j} = a_i/a_j
\end{align*}
for some $(a_1,\ldots,a_K)\in\CC^K$. The resulting channel
has entries $h_{i,j} = h_{j,j} a_i/a_j$, $j\in[i:K]$ and $i\in[1:K]$,
and zero for the remaining non-diagonal entries.
For these channels the sum-capacity is:
\begin{align}
\sum_{k=1}^{K} \log\left(1+\frac{|a_k|^2 \ |h_{k,k}/a_k|^2}{1+|a_k|^2 \sum_{j=k+1}^{K}|h_{j,j} /a_j|^2}\right).
\label{eq:successive decoding sumrate degdeg}
\end{align}
In the next subsection we relate the channels considered in Example~2
with the class of degraded channels studied in~\cite{josesriram:kifcmacbound:itw10}.

\subsection{Sum-capacity of non-Z channels}\label{sec:maingauss:nonZ}
The condition expressed by~\reff{eq:channel for succ dec}
is only sufficient for the achievability of~\reff{eq:successive decoding sumrate}.
In general, the expression in~\reff{eq:successive decoding sumrate}
is an upper bound to the sum-capacity of channels for which the upper triangular part of $\Hm$
can be expressed as in~\reff{eq:channel for succ dec} and the entries below the
main diagonal have any arbitrary value.
For such channels, the entries below the main diagonal might need to satisfy some extra
constraints (besides~\reff{eq:channel for succ dec}) in order for~\reff{eq:successive decoding sumrate} to be achievable.
The following discusses an example of such extra constraints.

\smallskip
The expression in~\reff{eq:successive decoding sumrate} suggests the following achievable
strategy for non-Z channels:  the $k$-th receiver first decodes users $1,\ldots,k-1$,
then strip them from its received signal, and finally decodes its intended message by
treating the signal of users $k+1,\ldots,K$ as noise; with this ``successive decoding
strategy'' the rate-tuplet $(r_1,\ldots,r_K)$ in~\reff{eq:succ dec rates}
is achievable if:
\begin{theorem}\label{eq:what can be achieved with successive decoding}
The sum-rate in~\reff{eq:successive decoding sumrate} is achievable
for a channel $\Hm$ such that the upper triangular part of $\Hm$
can be expressed as in~\reff{eq:channel for succ dec} and such that
the entries below the main diagonal satisfy:
\begin{align*}
&(r_1,\ldots,r_K) \in \bigcap_{k\in[1:K]}\bigcap_{\Sc_k\subseteq[1:k-1]}
\Big\{
(R_1,\ldots,R_K)\in\RR^{K}_{+}:
\\&
R_k + \sum_{j\in \Sc_k} R_j \leq \log\left(1+\frac{|h_{k,k}|^2+\sum_{j\in \Sc_k}|h_{k,j}|^2}{1+\sum_{i=k+1}^{K}|h_{k,i}|^2}\right)
\Big\}.
\end{align*}
for $(r_1,\ldots,r_K)$ defined in~\reff{eq:succ dec rates}. 
\end{theorem}
\begin{IEEEproof}
The proof follows from the previous discussion.
The achievable region in Th.\ref{eq:what can be achieved with successive decoding}
is the intersection of $K$ MAC regions where only the constraints that involve the
intended rate have been retained. 
\end{IEEEproof}

\medskip
As a corollary to Th.\ref{eq:what can be achieved with successive decoding} we have:
\begin{corollary}\label{cor:deg ch proof 1}
Degraded channels, that is, channels for which $\Hm$ has rank one, 
satisfy the assumptions of Th.\ref{eq:what can be achieved with successive decoding}.
\end{corollary}
\begin{IEEEproof}
Consider a $K$-IFC with unit rank channel matrix $\Hm=\av \bv^H$,
for some $K$-length column vectors $\av$ and $\bv$ (notice that these
channels have upper triangular part as in Example~2 with $b_k^* = h_{k,k}/a_k$). 
Without loss of generality, assume that 
the entries of the vector $\av$ satisfy $|a_1| \leq |a_2| ... \leq |a_K|$.
With this ordering the channel outputs from a Markov chain:
\begin{align*}
&X_{\rm eq}\defeq \sum_{k=1}^{X}  X_k b_k^* \to Y_K  \to Y_{K-1} \ldots \to Y_1,
\\
&Y_k\sim a_k X_{\rm eq} + Z_k, \ k\in[1:K].
\end{align*}
For this channel, clearly
the $k$-th decoder can decode all users with index $i<k$ without imposing 
any rate penalty to these users; thus sum-rate
in~\reff{eq:successive decoding sumrate} is achievable. 
\end{IEEEproof}

Corollary~\ref{cor:deg ch proof 1} offers a simple proof for the sum-capacity
result of~\cite{josesriram:kifcmacbound:itw10}; this implies that
Th.\ref{eq:what can be achieved with successive decoding}
generalizes the result of~\cite{josesriram:kifcmacbound:itw10}.

\medskip
Another proof for the converse part of Corollary~\ref{cor:deg ch proof 1} can be obtained 
by generalizing the bound for the degraded Gaussian 
2-IFC proposed by Sato in~\cite{sato:IFCoutit1978}. 
We have:
\begin{theorem}\label{th:out unitrank}
The capacity of the Gaussian $K$-IFC with channel matrix
$\Hm=\av \bv^H$, such that $|a_1| \leq |a_2| ... \leq |a_K|$,
is outer bounded by:
\begin{align}
R_k \leq \log\left(1+
\frac{ \beta_u \| \bv \|^2 |a_k|^2}{1+(\sum_{j=k+1}^{K}\beta_j \| \bv \|^2) |a_k|^2 }
\right),
\label{eq:extension KraTh2?}
\end{align}
for all $\beta_k\geq 0$, $k\in[1:K]$, such that 
$\sum_{k=1}^{K}\beta_k=1$. 
\end{theorem}
\begin{IEEEproof}
By letting the transmitters cooperate, the capacity of the
unit-rank $K$-IFC is outer bounded
by the capacity of a $K$-user degraded SISO Broadcast
Channel (BC) with input $X_{\rm eq} = \sum_{k=1}^K b_k^* X_k$, input power
constraint $\E[ X_{\rm eq}|^2 ] \leq \| \bv \|^2$, and outputs
$Y_k = a_k X_{\rm eq} + Z_k$, $k\in[1:K]$. 
The capacity of this degraded $K$-user BC is given by~\reff{eq:extension KraTh2?}~\cite{book:cover_thomas:it}.
\end{IEEEproof}

By using the outer bound in Th.\ref{th:out unitrank} we have the following 
very simple proof for the converse part of Corollary~\ref{cor:deg ch proof 1}:
by letting $\beta_u \| \bv \|^2 = |b_k|^2$ in~\reff{eq:extension KraTh2?}
we immediately obtain the upper bound in~\reff{eq:successive decoding sumrate}, which is
equivalent to~\reff{eq:successive decoding sumrate degdeg} with $b_k^* = h_{k,k}/a_k$.


\section{Conclusions}\label{sec:conc}
In this work we developed a framework to derive an
outer bound for the general memoryless interference channel
with an arbitrary number of source-destination pairs.  For the Gaussian channel,
we showed that the proposed bound gives the sum-capacity for certain
channels, including some Z-channels and degraded channels.

\appendix

{\bf Proof of~\reff{eq:th:out kra}.}
Consider a non-empty subset $\Sc$ of $[1:K]$ and let 
$\Sc^c$ be its complement in $[1:K]$.  Consider without 
loss of generality the permutation $\pi=(1,2,\ldots,|S|)$
(the others are obtained by relabeling the users).
We use the following conventions: for a set $\Sc$,
$W(\Sc) = \{W_i: i\in \Sc\}$. We have:
\begin{align*}
  &n \ \sum_{k=1}^{|\Sc|} (R_k - \epsilon_n) 
   \stackrel{\rm(a)}{=} \sum_{k=1}^{|\Sc|} I(X_k^n; Y_k^n)
\\&\stackrel{\rm(b)}{\leq} \sum_{k=1}^{|\Sc|} I(X_k^n; Y_k^n, \ Y_1^n,\ldots,Y_{k-1}^n,\ X_1^n,\ldots,X_{k-1}^n, \ X^n(\Sc^c))
\\&\stackrel{\rm(c)}{=}   \sum_{k=1}^{|\Sc|} I(X_k^n; Y_k^n, \ Y_1^n,\ldots,Y_{k-1}^n | X_1^n,\ldots,X_{k-1}^n, \ X^n(\Sc^c))
\\&\stackrel{\rm(d)}{=} \sum_{k=1}^{|\Sc|}\sum_{\ell=1}^{k}
 I(X^n_k; Y_\ell^n,|X^n_1,\ldots,X^n_{k-1}, \ X^n(\Sc^c), \ Y_1^n,\ldots,Y_{\ell-1}^n)
\\&\stackrel{\rm(e)}{=} \sum_{\ell=1}^{|\Sc|}\sum_{k=\ell}^{|\Sc|}
 I(X^n_k; Y_\ell^n,|X^n_1,Y_1^n,\ldots,X^n_{\ell-1},Y_{\ell-1}^n,\ X^n(\Sc^c),
\\&\qquad \qquad \qquad  \qquad \qquad  \qquad \qquad  \qquad
X^n_{\ell},\ldots,X^n_{k-1})
\\&\stackrel{\rm(f)}{=} \sum_{\ell=1}^{|\Sc|}
 I(X^n_\ell,\ldots,X^n_{|\Sc|}; Y_\ell^n,|X^n_1,Y_1^n,\ldots,X^n_{\ell-1},Y_{\ell-1}^n,\ X^n(\Sc^c))
\\&\stackrel{\rm(g)}{\leq}n \sum_{\ell=1}^{|\Sc|}
 I(X_\ell,\ldots,X_{|\Sc|}; Y_\ell|X_1,Y_1,\ldots,X_{\ell-1},Y_{\ell-1},X(\Sc^c),Q) 
\end{align*}
where the different (in)equalities follow from:
(a) Fano's inequality,
(b) non-negativity of mutual information (i.e., add side information at the receivers as in Fig.~\ref{fig:sideinfo}), 
(c) independence of messages (and thus of codewords),
(d) chain rule for mutual information,
(e) swap order of summation,
(f) chain rule for mutual information,
(g) ``conditioning reduces entropy'', memoryless property of the channel,
and by introducing a ``time sharing'' random variable $Q$ uniformly
distributed on $[1:n]$ and independent of everything else.

\medskip
{\bf Proof of~\reff{eq:th:out etw}.}
Consider a non-empty subset $\Sc$ of $[1:K]$,
\begin{align*}
  &\sum_{k\in\Sc}n(R_k-\epsilon_n)
   \leq \sum_{k\in\Sc} I(X_k^n; Y_k^n)
\\&\leq \sum_{k\in\Sc} I(X_k^n; Y_k^n, S_k^n)
\\&=    \sum_{k\in\Sc} H(S_k^n)
                      +H(Y_k^n|S_k^n)
                      -H(Y_k^n|X_k^n)
                      -H(S_k^n|X_k^n,Y_k^n)
\\&\stackrel{\rm(a)}{\leq} \sum_{k\in\Sc}
                       H(Y_k^n|S_k^n)
                      -H(S_k^n|Y_k^n,X_1^n,\ldots,X_K^n),
\end{align*}
where the inequality in (a) requires:
\[
\sum_{k\in\Sc} (H(S_k^n) - H(Y_k^n|X_k^n)
-  I(S_k^n; X^n_1,\ldots,X^n_K|X_k^n,Y_k^n)) \leq 0.
\]
A good candidate for the side information is as in Fig.~\ref{fig:sideinfo}
inspired by~\cite{etkin_tse_hua:withinonebit:subIt06}
\[
\{S_k, k\in \Sc \} =
\{Y_{\backslash k}, k\in \Sc\}, \quad \forall  \Sc \subseteq [1:K],
\]
where we defined
$Y_{\backslash k} \sim Y_k|_{X_k}$.


\section*{Acknowledgment}
This work was partially funded by NSF under award number 0643954.
The contents of this article are solely the responsibility of the authors and
do not necessarily represent the official views of the NSF.

\bibliographystyle{IEEEtran}
\bibliography{../career_biblio_v3}

\begin{thebibliography}{10}
\providecommand{\url}[1]{#1}
\csname url@samestyle\endcsname
\providecommand{\newblock}{\relax}
\providecommand{\bibinfo}[2]{#2}
\providecommand{\BIBentrySTDinterwordspacing}{\spaceskip=0pt\relax}
\providecommand{\BIBentryALTinterwordstretchfactor}{4}
\providecommand{\BIBentryALTinterwordspacing}{\spaceskip=\fontdimen2\font plus
\BIBentryALTinterwordstretchfactor\fontdimen3\font minus
  \fontdimen4\font\relax}
\providecommand{\BIBforeignlanguage}[2]{{%
\expandafter\ifx\csname l@#1\endcsname\relax
\typeout{** WARNING: IEEEtran.bst: No hyphenation pattern has been}%
\typeout{** loaded for the language `#1'. Using the pattern for}%
\typeout{** the default language instead.}%
\else
\language=\csname l@#1\endcsname
\fi
#2}}
\providecommand{\BIBdecl}{\relax}
\BIBdecl

\bibitem{sato:IFCoutit1978}
H.~Sato, ``On the capacity region of a discrete two-user channel for strong
  interference,'' in \emph{IEEE Trans.\ Inform.\ Theory}, vol. 24(3), May 1978,
  pp. 377--379.

\bibitem{costa_elgamal:strong:it1987}
M.~H.~M. Costa and A.~A.~E. Gamal, ``The capacity region of the discrete
  memoryless interference channel with strong interference,'' in \emph{IEEE
  Trans.\ Inform.\ Theory}, vol. 33(5), Sept 1987, pp. 710--711.

\bibitem{carleial:ifcstrong:it1975}
A.~B. Carleial, ``A case where interference does not reduce capacity,'' in
  \emph{IEEE Trans.\ Inform.\ Theory}, vol. 21(5), Sept 1975, pp. 569--570.

\bibitem{costa_aelgamal:it1982}
A.~A.~E. Gamal and M.~H.~M. Costa, ``The capacity region of a class of
  deterministic interference channels,'' in \emph{IEEE Trans.\ Inform.\
  Theory}, vol. 28(2), March 1982, pp. 343--346.

\bibitem{liu_ulukus:degifc}
N.~Liu and S.~Ulukus, ``The capacity region of a class of discrete degraded
  interference channels,'' in \emph{IEEE Trans.\ Inform.\ Theory}, vol. 54(9),
  Sept 2008, pp. 4372--4378.

\bibitem{benzel:it1979}
R.~Benzel, ``The capacity region of a class of discrete additive degraded
  interference channels,'' \emph{IEEE Trans.\ Inform.\ Theory}, vol. IT-25, pp.
  228--231, Mar 1979.

\bibitem{Han_Kobayashi:it1981}
T.~S. Han and K.~Kobayashi, ``A new achievable rate region for the interference
  channel,'' in \emph{IEEE Trans.\ Inform.\ Theory}, vol. 27(1), Jan 1981, pp.
  49 --60.

\bibitem{sato:IFCoutit1977}
H.~Sato, ``Two-user communication channels,'' in \emph{IEEE Trans.\ Inform.\
  Theory}, vol. 23(3), 1977, pp. 295 -- 304.

\bibitem{carleial:IFCit1983}
A.~B. Carleial, ``Outer bounds on the capacity of interference channels,'' in
  \emph{IEEE Trans.\ Inform.\ Theory}, vol.~29, no.~4, July 1983, pp. 602--606.

\bibitem{kramer:it2004}
G.~Kramer, ``Outer bounds on the capacity of gaussian interference channels,''
  in \emph{IEEE Trans.\ Inform.\ Theory}, vol.~50, no.~3, Jan 2004, pp.
  581--586.

\bibitem{sato:IFCstrong:it1981}
H.~Sato, ``The capacity of the gaussian interference channel under strong
  interference,'' in \emph{IEEE Trans.\ Inform.\ Theory}, vol. 27(6), Nov 1981,
  pp. 786--788.

\bibitem{costa:awgnifc:it1985}
M.~H.~M. Costa, ``On the gaussian interference channel,'' in \emph{IEEE Trans.\
  Inform.\ Theory}, vol. 31(5), Sept 1985, pp. 607--615.

\bibitem{tuninettiweng:isit2008}
D.~Tuninetti and Y.~Weng, ``On gaussian mixed interference channels,'' in
  \emph{IEEE International Symposium Information Theory}, Toronto, Canada, July
  2008.

\bibitem{canadanoisy}
A.~S. Motahari and A.~K. Khandani, ``Capacity bounds for the gaussian
  interference channel,'' \emph{IEEE Trans.\ Inform.\ Theory}, vol.~55, no.~2,
  pp. 620--643, 2009.

\bibitem{sason:it2004}
I.~Sason, ``On achievable rate regions for the gaussian interference channel,''
  in \emph{IEEE Trans.\ Inform.\ Theory}, vol. 50(6), June 2004, pp.
  1345--1356.

\bibitem{biaonoisy2008:IT09}
X.~Shang, G.~Kramer, and B.~Chen;, ``A new outer bound and the
  noisy-interference sumÐrate capacity for gaussian interference channels,'' in
  \emph{IEEE Trans.\ Inform.\ Theory}, vol. 55(2), Feb 2009, pp. 689 -- 699.

\bibitem{venunoisy2008:IT09}
V.~S. Annapureddy and V.~V. Veeravalli, ``Gaussian interference networks: Sum
  capacity in the low-interference regime and new outer bounds on the capacity
  region,'' in \emph{IEEE Trans.\ Inform.\ Theory}, vol. 55(7), July 2009, pp.
  3032 -- 3050.

\bibitem{etkin_tse_hua:withinonebit:subIt06}
R.~H. Etkin, D.~N.~C. Tse, and H.~Wang, ``Gaussian interference channel
  capacity to within one bit,'' \emph{IEEE Trans.\ Inform.\ Theory}, vol.~54,
  no.~12, pp. 5534 -- 5562, 2008.

\bibitem{EtkinISIT2009}
R.~Etkin, ``New sum-rate upper bound for the two-user gaussian interference
  channel,'' \emph{Information Theory, 2009. ISIT 2009. IEEE International
  Symposium on}, pp. 2582 -- 2586, 2009.

\bibitem{tuniITA11:G3IFCsumrate}
D.~Tuninetti, ``A new sum-rate outer bound for interference channels with three
  source-destination pairs,'' \emph{Proceedings of the Information Theory and
  Applications Workshop, 2011, San Diego, CA USA}, Feb 2011.

\bibitem{Cadambe-Jafar:arXiv:0707.0323v2}
V.~R. Cadambe and S.~A. Jafar, ``Interference alignment and the degrees of
  freedom for the k user interference channel,'' \emph{IEEE Trans.\ Inform.\
  Theory}, vol.~54, no.~8, pp. 3425--3441, Aug 2008.

\bibitem{Nazer:2007}
B.~Nazer and M.~Gastpar, ``Computation over multiple-access channels,''
  \emph{IEEE Trans.\ Inform.\ Theory}, vol.~53, no.~10, pp. 3498--3516, Oct.
  2007.

\bibitem{Bresler:1toM_Mto1:allerton07}
G.~Bresler, A.~Parekh, , and D.~Tse, ``The approximate capacity of the
  many-to-one and one-to-many gaussian interference channel, arxiv:0809.3554,''
  \emph{Proceedings of the Allerton Conference}, Sept 2007.

\bibitem{Sridharan:layredlattice:allerton08}
S.~Sridharan, A.~Jafarian, S.~Vishwanath, S.~Jafar, and S.~Shamai, ``A layered
  lattice coding scheme for a class of three user gaussan interference
  channels,'' \emph{Communication, Control, and Computing, 2008 46th Annual
  Allerton Conference on}, pp. 531 -- 538, Sept 2008.

\bibitem{cadambe:notseparable:it09}
V.~R. Cadambe and S.~A. Jafar, ``Parallel gaussian interference channels are
  not always separable,'' \emph{IEEE Trans.\ Inform.\ Theory}, vol. 55(9), pp.
  3983--3990, Sept 2009.

\bibitem{josesriram:kifcmacbound:itw10}
J.~Jose and S.~Vishwanath, ``Sum capacity of k user gaussian degraded
  interference channels,'' \emph{Proc. IEEE Info. Theory Workshop (ITW),
  Dublin, Ireland.}, Sept 2010.

\bibitem{ZhouYu:cyclicsym:CISS10}
L.~Zhou and W.~Yu, ``On the symmetric capacity of the k-user symmetric cyclic
  gaussian interference channel,'' \emph{Proceedings of Conference on
  Information Science and Systems (CISS), Princeton, NJ}, March 2010.

\bibitem{Bandemer:cyclicsym:isit09}
B.~Bandemer, G.~Vazquez-Vilar, and A.~E. Gamal, ``On the sum capacity of a
  class of cyclically symmetric deterministic interference channels,''
  \emph{Proc. IEEE International Symposium on Information Theory, Seoul,
  Korea}, p. 2622Ð2626, July 2009.

\bibitem{segzin_3ifc:arXiv:1010.4911}
A.~Chaaban and A.~Sezgin, ``The capacity region of the 3-user gaussian
  interference channel with mixed strong-very strong interference,
  arxiv:1010.4911,'' \emph{Submitted to ISIT 2011}, 2010.

\bibitem{Jovicic_Wang_Viswanath:itw2007}
J.~A., H.~Wang, and V.~P., ``On network interference management,'' pp. 307 --
  312, Sept 2007.

\bibitem{Sridharan:verystrongKifc:globecom08}
S.~Sridharan, A.~Jafarian, S.~Vishwanath, and S.~A. Jafar, ``Capacity of
  symmetric k-user gaussian very strong interference channels,'' \emph{Globecom
  2008}, 2008.

\bibitem{jafar:dofKifc:it10}
S.~A. Jafar and S.~Vishwanath, ``Generalized degrees of freedom of the
  symmetric gaussian k user interference channel,'' \emph{IEEE Trans.\ Inform.\
  Theory}, vol. 56(7), pp. 3297--3303, July 2010.

\bibitem{EtkinOrdentlichISIT2009}
R.~Etkin and E.~Ordentlich, ``On the degrees-of-freedom of the k-user gaussian
  interference channel,'' \emph{Information Theory, 2009. ISIT 2009. IEEE
  International Symposium on}, pp. 1919 -- 1923, 2009.

\bibitem{jose:icc2011}
J.~Jose, N.~Prasad, M.~Khojastepour, and S.~Rangarajan, ``On robust
  weighted-sum rate maximization in mimo interference networks,'' \emph{Proc.
  IEEE ICC 2011, Kyoto, Japan}, June 2011.

\bibitem{biaonoisy2008:ISIT08}
X.~Shang, G.~Kramer, and B.~Chen, ``New outer bounds on the capacity of
  gaussian interference networks,'' in \emph{IEEE Int. Symp. Inform. Theory,
  Toronto, Canada}, July 2008, pp. 245 -- 249.

\bibitem{book:cover_thomas:it}
T.~Cover and J.~Thomas, \emph{Elements of information theory}.\hskip 1em plus
  0.5em minus 0.4em\relax New York: Wiley, 1991.

\bibitem{viswanathextrema}
T.~Liu and P.~Viswanath, ``An extremal inequality motivated by multiterminal
  information theoretic problems,'' \emph{IEEE Trans.\ Inform.\ Theory},
  vol.~53, no.~5, pp. 1839 -- 1851, 2007.

\bibitem{RTDjournal1}
S.~Rini, D.~Tuninetti, and N.~Devroye, ``New inner and outer bounds for the
  discrete memoryless cognitive channel and some capacity results,''
  \emph{Information Theory, IEEE Transactions on}, 2010, in print, preprint at
  arXiv:1003.4328.

\bibitem{rini2010CIFC+CR:ITWDublin}
------, ``Outer bounds for the interference channel with a cognitive relay,''
  \emph{Proc. IEEE Information Theory Workshop (ITW), Dublin, Ireland (preprint
  at arXiv:1004.4944)}, Sep 2010.

\bibitem{tuniISIT09out}
S.~Yang and D.~Tuninetti, ``A new sum-rate outer bound for gaussian
  interference channels with generalized feedback,'' in \emph{Proceedings of
  2009 IEEE International Symposium on Information Theory (ISIT 2009), Seoul,
  South Korea}, June 2009.

\bibitem{tuniITA10out}
D.~Tuninetti, ``An outer bound region for interference channels with
  generalized feedback,'' \emph{Proceedings of the Information Theory and
  Applications Workshop, 2010, San Diego, CA USA}, Jan 2010.

\bibitem{vinodSC2009}
V.~Prabhakaran and P.~Viswanath, ``Interference channels with source
  cooperation,'' \emph{submitted to IEEE Trans. Info. Theory in May 2009, Arxiv
  preprint arXiv:0905.3109v1}, 2009.

\end{thebibliography}

\end{document}